\documentclass[]{spie}  
\usepackage[]{graphicx}
\title{Development status of the LAUE project} 
\author{F.~Frontera\supit{a,b}, E.~Virgilli\supit{a}, V.~Liccardo\supit{a,c}, V.~Valsan\supit{a,c}, V.~Carassiti\supit{d}, S.~Chiozzi\supit{d}, F.~Evangelisti\supit{d}, S.~Squerzanti\supit{d}, M.~Statera\supit{a}, V.~Guidi\supit{a}, C.~Ferrari\supit{e}, R.~A.~Zappettini\supit{e}, E.~Caroli\supit{b}, N.~Auricchio\supit{b}, S.~Silvestri\supit{b}, R.~Camattari\supit{a}, F.~Cassese\supit{f}, L.~Recanatesi\supit{f}, M.~Pecora\supit{g}, S.~Mottini\supit{h}, B.~Negri\supit{k}
\skiplinehalf
\supit{a} \small\textit{Physics Department, University of Ferrara - Via Saragat, 1, 44100 Ferrara - Italy};\\
\supit{b} \small\textit{INAF-IASF Bologna, via P.~Gobetti 101, Bologna - Italy};\\
\supit{c}\textit{Universit\'e de Nice Sophia Antipolis, Nice, Cedex 2, Grand Chateau Parc Valrose - France;}\\
\supit{d} \small\textit{INFN, Sezione di Ferrara, via Saragat 1, 44100 Ferrara - Italy};\\
\supit{e} \small\textit{CNR, IMEM, Parco Area delle Scienze 37/A - 43124 Parma - Italy};\\
\supit{f} \small\textit{DTM, Modena, Via Tacito, I-41100 Modena - Italy};\\
\supit{g} \small\textit{Thales Alenia Space--Italy, Milan - Italy};\\
\supit{h} \small\textit{Thales Alenia Space--Italy, Turin - Italy};\\
\supit{k} \small\textit{ASI, Agenzia Spaziale Italiana, Viale Liegi 26, I-00198 Roma - Italy}.
}

\authorinfo{E-mail to: frontera@fe.infn.it}
\begin{document} 
\maketitle 

\begin{abstract}
We present the status of LAUE, a project  supported by the Italian Space Agency (ASI),  and devoted to 
develop Laue lenses with long focal length (from 10--15 meters up to 100 meters), for hard X--/soft gamma--ray astronomy 
(80-600 keV). Thanks to their focusing capability, the design goal is to improve the sensitivity of the 
current instrumention in the above energy band by 2 orders of magnitude, down to a 
few times $10^{-8}$~photons/(cm$^2$~s~keV).
\end{abstract}


\section{INTRODUCTION}
\label{sec:intro}

Many astrophysical issues are expected to be solved with focusing  telescopes that cover the soft 
gamma--ray band (beyond 80/100 keV), if they can reach a sensitivity a factor from 10 to 100 better 
than that of the current instrumentation (e.g., IBIS instrument aboard INTEGRAL\cite{Ubertini03}~). 
A discussion of the achievable scientific obiectives has been extensively discussed in 
more occasions (see, e.g., Ref.~[\citenum{Frontera10}]). 

Among the many astrophysical issues that could be settled, we like to mention one of the most elusive questions 
since many years: the origin of positrons in the Galactic Center (CG) region. 
A  SPI/INTEGRAL all-sky map of the galactic e$^-$/e$^+$ annihilation radiation at 511 keV
shows an asymmetric distribution around the GC\cite{Weidenspointner08}~. The authors of the paper
give a possible interpretation in terms of superposition of the discrete positron annihilation radiation sources. 
These sources should  belong of a 
special class of Low Mass X--ray Binaries (called hard LMXBs) located in the CG region, that are strong 
emitters of hard X--rays ($>$20 keV). However other possible interpretations have been advanced by other authors, 
such as the presence of an extended region of antimatter or dark matter, or the existence of a source of radioactive 
elements, like $^{26}$Al, $^{56}$Co, $^{44}$Ti, or the presence of a gamma--ray source (e.g., gamma--ray pulsar).    
Only much more sensitive observations can settle the origin of  
the annihilation line from the GC region.

We have already reported on the LAUE project last year\cite{Virgilli11b}~. It is the follow-up of an initial 
development project (HAXTEL), also supported by ASI, 
that allowed to implement a  concept of assembling technique for Laue lenses\cite{Frontera08a,Virgilli11a}~, 
to significantly shorten the time needed to build a lens. On the basis of the tests performed 
on two lens prototypes built according to the HAXTEL technology, we found that the implemented technique is 
satisfactory only for short focal-length Laue lenses ($<$10 m). 

Given the sensitivity and the energy broad-band to be covered with the first generation of these
conceptually new focusing telescopes, focal lengths longer than 10 meters are required. These reasons led 
the birth of a new project named LAUE, that was preliminary discussed in Ref.~[\citenum{Virgilli11b}].
Now the project is definitive and it can be better described.

\newpage

\section{The LAUE project: a focusing lens for soft  gamma--rays} 
\label{s:laue}

\begin{figure}[!h]
\begin{center}
\includegraphics[width=0.30\textwidth]{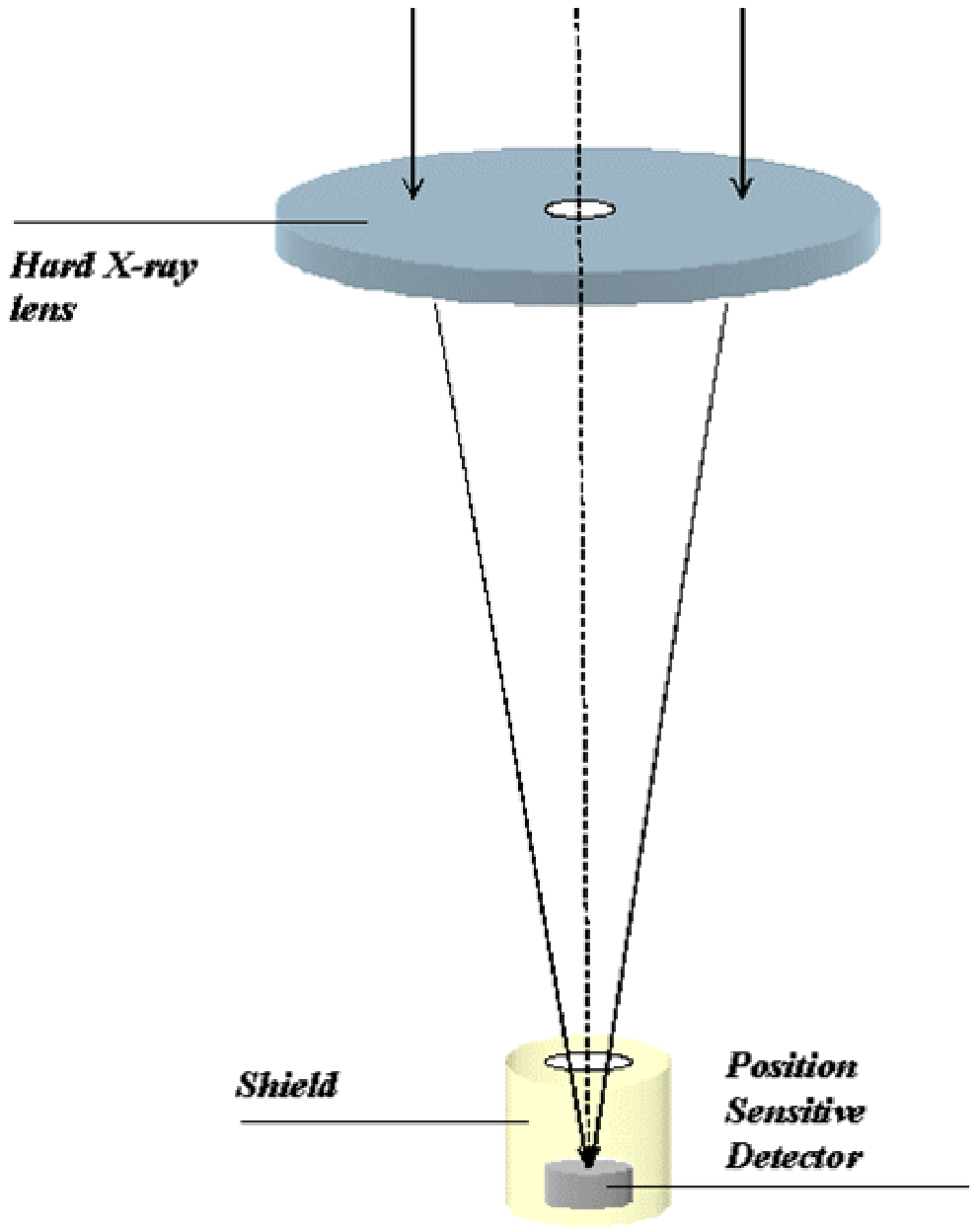}
\includegraphics[angle=0,width=0.27\textwidth]{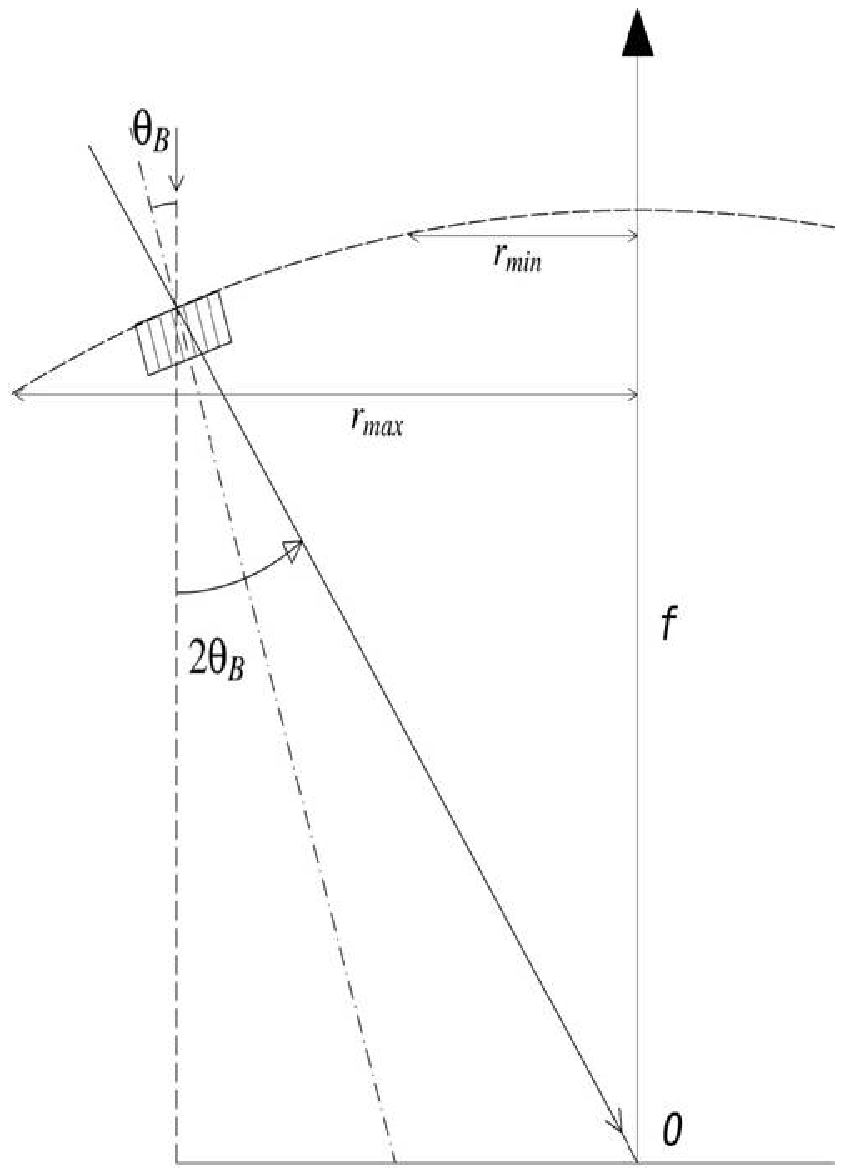}
\end{center}
\caption{{\em Left~panel}: sketch of a Laue lens.
{\em Right~panel}: orientation of crystals in a Laue lens.}
\label{f:lens-design} 
\end{figure}

The sketch of a Laue Lens for space astrophysics and the required orientation of the crystals in the lens 
are shown in Figure~\ref{f:lens-design}. An extensive  review on Laue lenses can be found in Ref.~[\citenum{Frontera10}].
   
%

The main goal of the LAUE project is to develop an advanced technology for building a sensitive Laue lens 
with broad energy band ($>$80/100 keV) and long focal length (up to 100 m) for space astrophysics but 
the project also faces a still long standing and difficult issue: the development of suitable crystals for 
gamma--ray lenses.

\begin{figure}
\begin{center}
\includegraphics[width=0.4\textwidth]{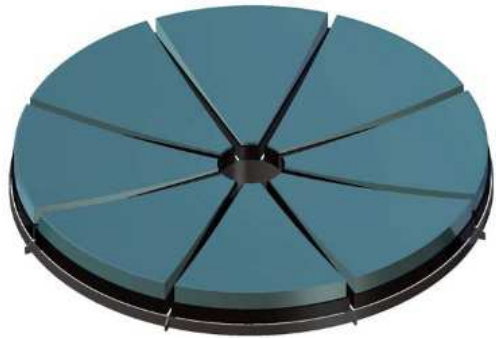}
\end{center}
\caption{View of a space lens made of petals. The building of one of these petals
is one of the goals of the LAUE project.}
\label{f:space-lens}
\end{figure}

From the experience gained with the HAXTEL project, we have adopted the following approach for positioning the 
crystals on a lens frame. Under the control of a gamma--ray beam, each crystal tile is properly positioned when 
it focuses the 
reflected photon beam to the lens focus. Once the correct orientation is found, each crystal tile is fixed to 
the lens frame, which is kept fixed during the lens assembling process. This implies that the beam has to be 
continuously translated to assemble the crystal tiles.
This approach is expected to minimize the orientation errors. 
Given the large size (diameter larger than 1 m) of space lenses, the only feasible way to build such lenses, 
is that of subdividing them in a number of lens petals. The feasibility study of a similar approach for building 
a space lens is being performed within the LAUE project
(see Figure~\ref{f:space-lens}). One of the lens petals, with 20 m focal length and filled with about 
300 crystals tiles, will be built as a result of the LAUE project.  

%
%

\section{Apparatus for lens assembling}
\label{s:apparatus}
 
\begin{figure}[!h]
\begin{center}
\includegraphics[width=0.75\textwidth]{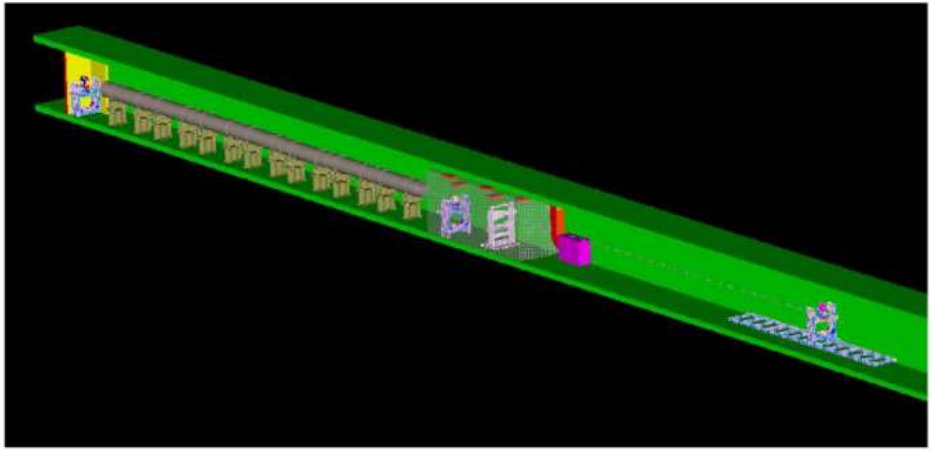}
\end{center}
\caption{Drawing of the LAUE apparatus located in the tunnel of the LARIX facility. It is visible
the X-.ray source surrounded by the protection lead wall (on the left), the beam line, the clean room with controled temperature and environment in which 
the lens will be assembled (at the center), and the focal plane detector holder mounted on a rail.}
\label{f:larix} 
\end{figure}

A lens assembling  apparatus is being installed 
in a tunnel of the LARge Italian X--ray Facility (LARIX) 
of the Physics Department of the University of Ferrara. Its drawing is shown in Figure~\ref{f:larix}. 
The two enters of the tunnel are properly shielded from the X/gamma radiation. On one side, close to the 
B section of the LARIX facility, an \texttt{L}-shaped lead wall surrounding the source is set. On the opposite side, beyond
the detectors, a motorized lead door 65 mm thick has been already installed. This guarantees a free use of the 
laboratories located outside the tunnel: LARIX~A at the beginning of the tunnel, and LARIX~B at its end. 
The former section will be used as remote control room for managing the LAUE apparatus and the petal 
lens assembling. 

%

The main subsystems of the apparatus include a gamma--ray generator, a beam-line, a clean room, with temperature and 
humidity control, in which the lens petal will be assembled, and a focal plane detection system, with its holder mounted 
on a rail that can be moved back and forth.

The clean room hosts the following subsystems: 1) a slit collimator for controlling  the size of the gamma--ray 
beam impinging on the crystal tile, 2) the lens frame, 3) a multi-axis positioning system for correctly 
positioning and orienting crystal tiles on the lens, 4) an automated system for fixing crystal tiles to the frame.
In the following subsections details about the main apparatus components are presented.

\subsection{Gamma--ray source system}

The gamma--ray source system includes a portable betatron and an hard X--ray generator. 
The betatron has a maximum electron energy that can be regulated between 1 and 2.5 MeV, with a 
Tungsten target on which the accelerated electrons impinge for producing bremsstrahlung radiation. 
The maximum betatron power is 310 W, with a gamma--ray focal spot less than 0.2 $\times$ 3 mm.

%
 \begin{figure}
\begin{center}
\includegraphics[width=0.9\textwidth]{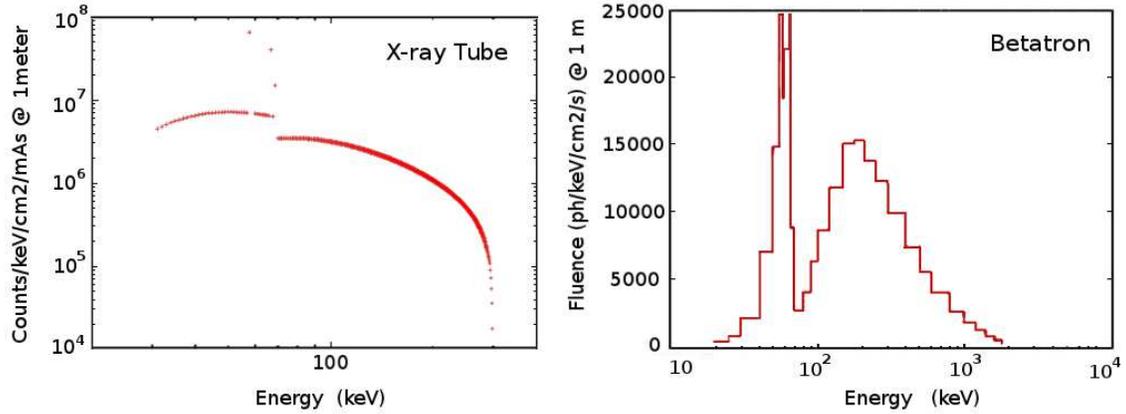}
\end{center}
\caption{Expected photon spectrum (in photons~keV$^{-1}$~cm$^{-2}$~s$^{-1}$) at 1 m distance of the X--ray tube ($left$)
and from the betatron ($right$).}
\label{f:sources} 
\end{figure} 

The hard X--ray generator has a maximum voltage of 320 kV, an X--ray tube with a tungsten anode and a fine 
focus of 0.2 mm radius. The maximum releasable power is $\sim$ 1800 W. The expected photon spectrum from 
the hard X--ray generator and for the betatron are shown in Figure~\ref{f:sources}.

The X--ray tube, providing a higher fluence than that of the betatron, will be used for the lens building phase, given 
that a higher fluence imply a shorter integration time for the tiles allignment stage. On the other hand, the betatron 
source can be used for the performence tests and, more important, will allow to build Laue lenses focusing higher energies
than that designed lens petal.

The gamma--ray source holder can be remotely translated along two perpendicular axes y and z 
(vertical) (the x axis is directed along the tunnel direction) and can be tilted around the z axis.

\subsection{Beam-line}

\begin{figure}
\begin{center}
\includegraphics[width=0.75\textwidth]{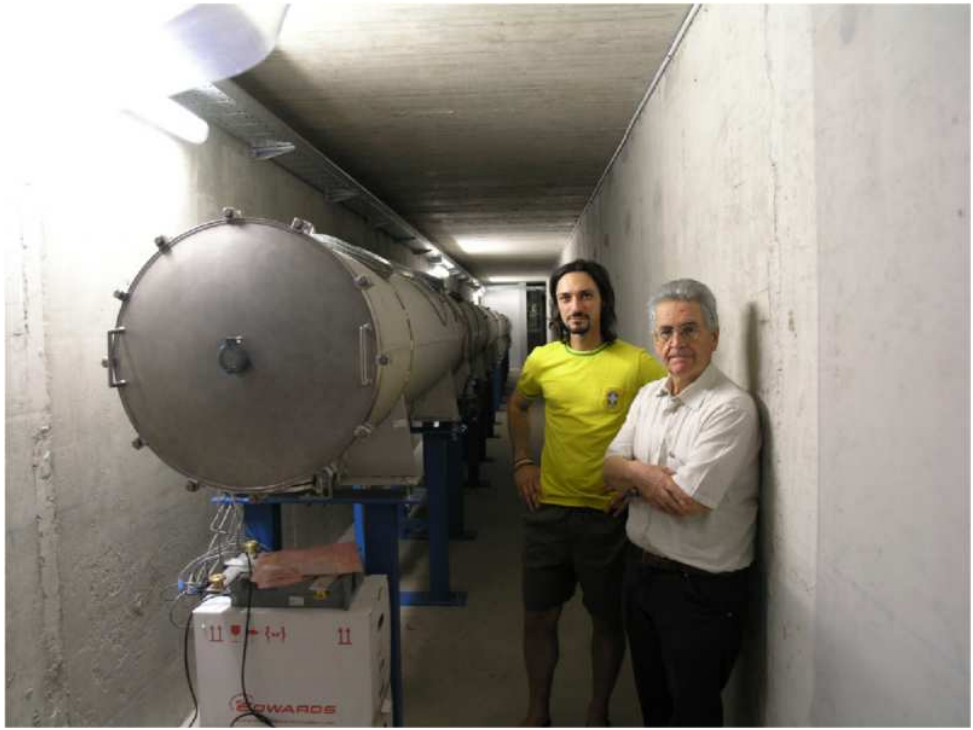}
\end{center}
\caption{A view of the 21~m beam-line installed in the LARIX tunnel.}
\label{f:beamline} 
\end{figure}

The photons coming from the gamma--ray source, after an initial collimation, enter in a 
21~m long beam-line. It is made of seven 3~m long tubes of stainless steel with an internal diameter of 60~cm. 
A set of vacuum pumps guarantees a vacuum of 1 mbar inside the tubes. The X--ray entrance and exit windows of the 
beam-line are made of carbon fiber 3 mm thick. The prospect is to extend the beam-line will be extended up to 70 m.
A view of the already installed beam-line is shown in Figure~\ref{f:beamline}.

%

\subsection{Clean-room and the main apparatus components inside}

\begin{figure}[!t]
\begin{center}
\includegraphics[width=0.75\textwidth]{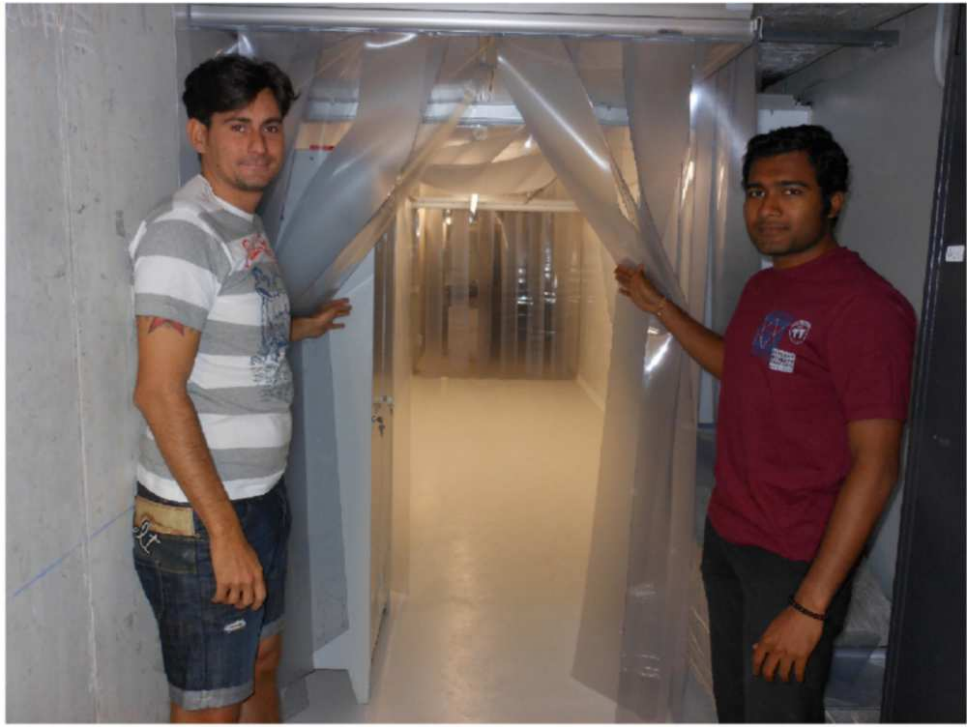}
\end{center}
\caption{A view of the entrance in the clean-room already installed in the LARIX tunnel.}
\label{f:cleanroom} 
\end{figure}

Once the gamma--ray beam exits from the 
beam-line, it enters in a clean-room (class better than 10$^5$, US FED STD 209E Cleanroom Standards). The clean-room is endowed with a thermal control 
(within $1\,^{\circ}{\rm C}$ accuracy) and an hygrometric control (relative humidity $\Phi$ = 60\% within 
an error of 10\%). A view of the entrance in the clean-room is shown in Figure~\ref{f:cleanroom}.
The main components inside the clean-room are the following.

%
%

\begin{figure}[!t]
\begin{center}
\includegraphics[width=0.75\textwidth]{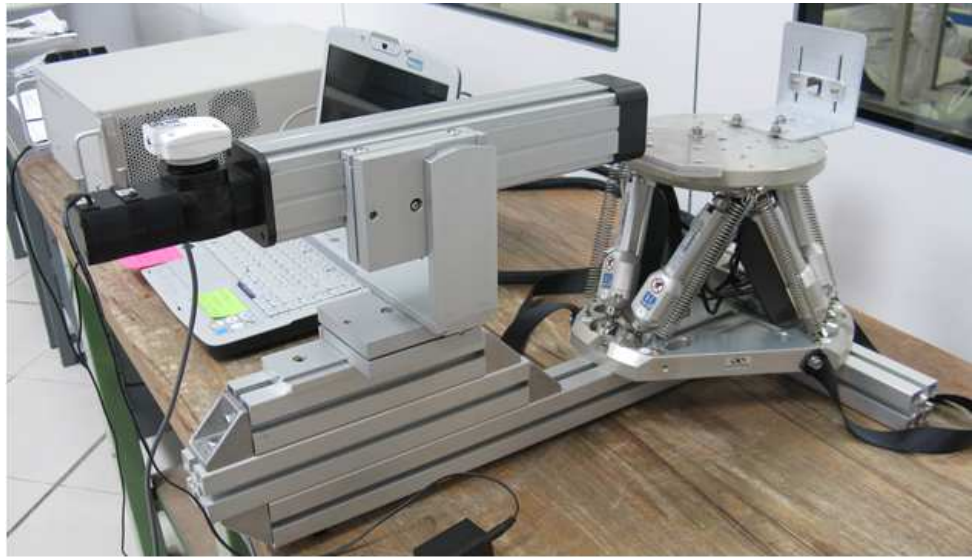}
\end{center}
\caption{A view of the esapod under test.}
\label{f:esapod} 
\end{figure}

\begin{itemize}
\item {\bf Slit collimator}

A slit collimator, made of four independent blades of Tungsten 20 mm thick, with variable aperture along two 
ortogonal directions allows to regulate, from the remote control room, the size of the gamma--ray beam 
coming from the beam-line. The beam size establishes the beam divergence.
In addition, the slit collimator can be accurately translated along the y and z  axes and 
rotated around the z and x axes.
The collimator translation follows  the source translation, in order to have the direction of 
the gamma--ray beam always parallel to the lens axis.

\item {\bf Frame of the lens petal}

The frame of the lens petal, on which the crystal tiles will be pasted, is made of carbon fiber. 
A view of the petal frame breadboard is shown in Figure~\ref{f:frame}.

%
%
\begin{figure}[!t]
\begin{center}
\includegraphics[width=0.75\textwidth]{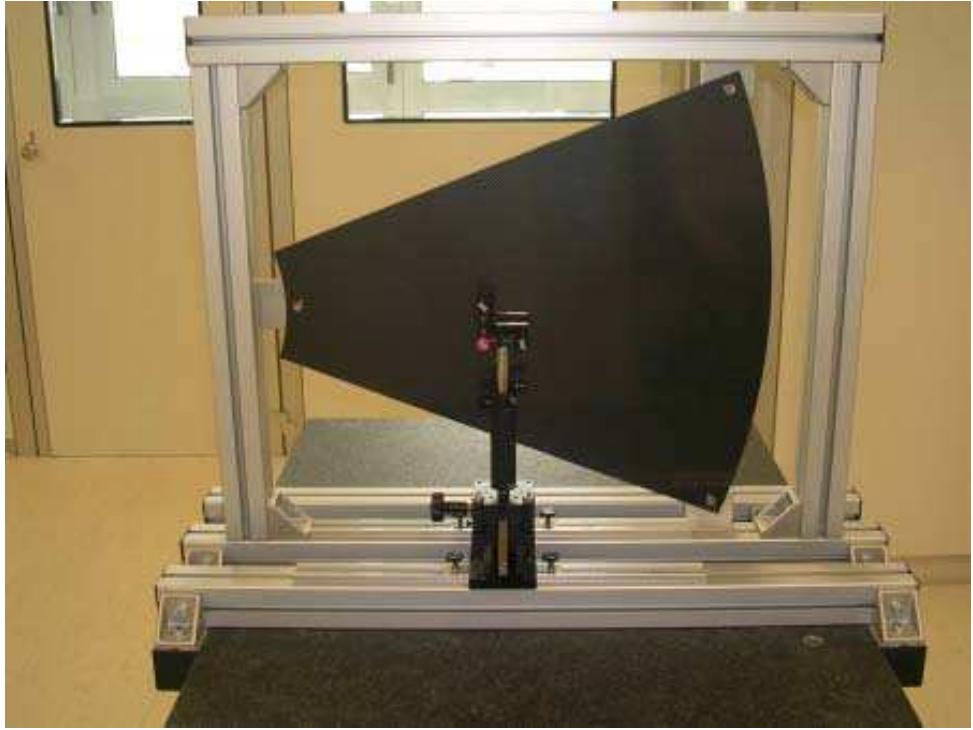}
\end{center}
\caption{A view of the breadboard of the petal frame.}
\label{f:frame} 
\end{figure} 

\item {\bf Crystal positioning system}

The positioning of each crystal tile on the petal frame is performed by using a mechanical system
 made of a coarse positioner and a fine positioner. The coarse positioner is made of 
two perpendicular translation stages, while the fine positioner, also called esapod, for its 
6 degrees of freedom, allows to accurately control the crystal tilting around two perpendicular axes   
and its motion  toward the frame.  A view of the esapod under test is shown in 
Figure~\ref{f:esapod}. 
\end{itemize}

%
%

\section{Focal plane detection system}

A gamma--ray detector system allows to establish the correct orientation of each crystal tile.
It is made of two focal plane detectors: 1) an X--ray imaging detector with a useful area 20$\times$20~cm$^2$, with 
a spatial resolution 200 $\mu$m and an operational energy range from 20 keV to 15 MeV, 2) a cooled HPGe spectrometer 
with 2.5 cm diameter and an energy resolution about 500 eV at 122 keV.

\begin{figure}[!t]
\begin{center}
\includegraphics[width=0.75\textwidth]{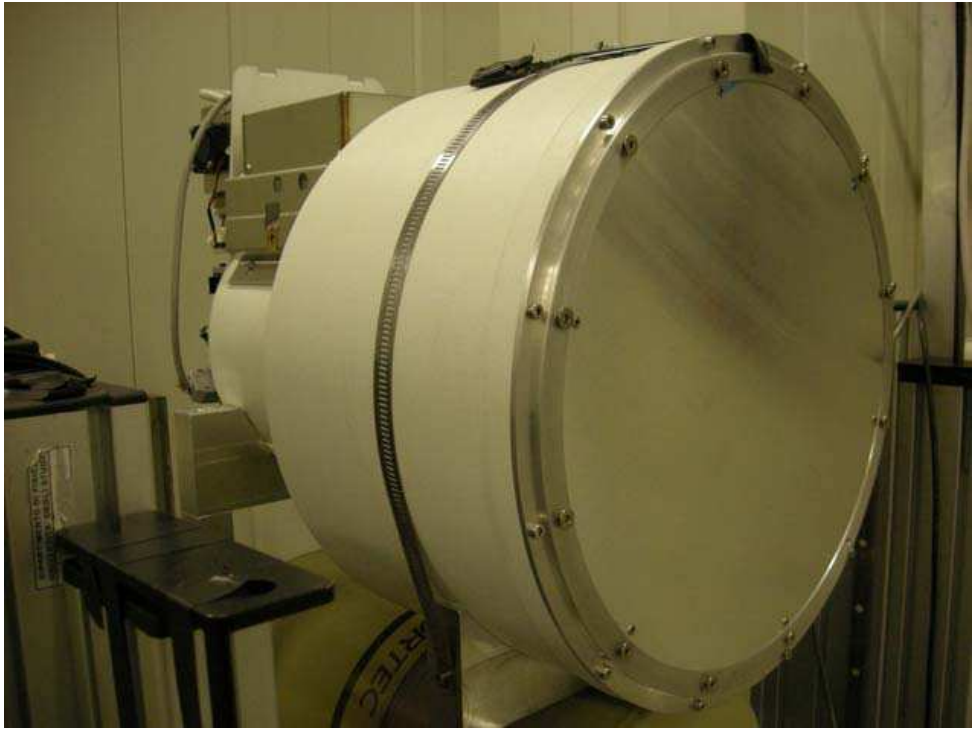}
\end{center}
\caption{A view of an imaging detector similar to that used for the LAUE project.}
\label{f:detector} 
\end{figure}

In addition to the possibility to be moved back and forth, the detector system can be translated along y and 
z axes and can be rotated along the x axis. A view of the imaging detector is shown in Figure~\ref{f:detector}.

%
%

\subsection{Managing of the lens assembling system}

As discussed above, the LAUE apparatus is located in the LARIX tunnel. It is controlled from a remote console, 
located in the section A of the LARIX facility that manages
all the required functions for calibrating the facility and for assembling a lens petal. 
The main functions that can be managed from the remote console are the following:
\begin{itemize}
\item
to set up the gamma--ray beam features (e.g., intensity, size) and translate its axis within the 
beam-line tube cross-section; 
\item 
to position the slit collimator such that the beam axis is kept parallel to the lens axis;
\item
to properly orient the crystal tile before it can fixed to the petal frame; 
\item 
to acquire spectra and images from the focal plane detectors;
\item
to determine the position error of the reflected beam with respect to the lens focus.
\item
to send the position error to the esapod for a further crystal orienting step.
\end{itemize}

\section{Crystals adopted for the lens petal} 

After a development phase, the crystals that will be used for lens petal will be a combination of
bent perfect crystals of Ge(111) and bent mosaic crystals of  GaAs(111). The GaAs mosaicity is about 20 arcsec. 
The advantage of bent crystals is their better focusing properties (see Ref.~[\citenum{Frontera10}]).

The crystal cross section has been chosen to be 30 $\times$ 10 mm$^{2}$, with the longer side 
radially placed on the lens frame. The main advantages of the rectangular shape, 
together with the radial disposition, concerns the focusing effect provided by bent crystals, 
which only acts in the radial direction. In such a way, a shorter tangential dimension provides 
a smaller defocusing factor, being proportional to the tile size.  
On the other hand, a bigger radial dimension allows to decrease the total number of crystals, reducing
the error budget potentially caused by each crystal misalignment contribution. 

The thickness $t$ of the crystal tiles is 2 mm for both crystal materials. The thickness value is imposed
by the current status of the bending technology adopted for bending Ge\cite{Guidi11} and 
GaAs\cite{Buffagni12}~.  The expected performance of the adopted crystals and the experimental reflectivity 
results obtained from the laboratory tests performed in tha LARIX facility are discussed in
two papers, one by Valsan et al.\cite{Valsan12} and the other by Liccardo et al.\cite{Liccardo12}~. 
Both will be published, along with our paper, in the same issue of the SPIE Proceedings.

\section{Conclusions}

From the already gained experience on Laue lenses (see, e.g., [\citenum{Frontera08a,Virgilli11a}]), we have started
a new project, LAUE, supported by the Italian Space Agency (ASI), devoted to the development of
an advanced lens assembling technology that we expect it eventually will allow to accurately build Laue
lenses for space astrophysics. The expected accuracy in the lens assembling would allow to build
lenses even with very long focal lengths (up to 100 m), a goal never achieved so far.
The LAUE apparatus, described in this paper,  is being installed in the LARIX facility of the University of Ferrara. 
We expect in a few months to start the assembling activity of a lens petal, made of about 300 bent crystals of 
Ge(111) and GaAs(111). Results on the built lens petal prototype are expected to be reported in the next year 
SPIE plenary conference.

\acknowledgments     

The LAUE project is the result of big efforts made by a large number of organizations and people.
We would like to thank all of them. We also acknowledge the ASI Italian Space Agency for its support 
to the LAUE project under contract I/068/09/0.

Vincenzo Liccardo and Vineeth Valsan are supported by the Erasmus Mundus Joint Doctorate Program by Grant Number
2010-1816 from the EACEA of the European Commission.

\bibliography{lens_biblio}
\bibliographystyle{spiebib}

\end{document}